\documentclass[twocolumn,times]{aastex62}
\usepackage{starType}

\newcommand*{\Qimp}{\ensuremath{Q_{\mathrm{imp}}}}
\newcommand*{\ntr}{\ensuremath{n_{\mathrm{t}}}}
\newcommand*{\nsat}{\ensuremath{n_{\mathrm{s}}}}

\graphicspath{{./}{figures/}}

\shorttitle{Crust Cooling Models are Insensitive to the Crust-Core Transition Pressure for Realistic EoS}
\shortauthors{Lalit, Meisel, \& Brown}

\begin{document}

\title{Crust Cooling Models are Insensitive to the Crust-Core Transition Pressure for Realistic Equations of State}

\author[0000-0001-7758-492X]{Sudhanva Lalit}
\altaffiliation{Affiliated with the Joint Institute for Nuclear Astrophysics—\\Center for the Evolution of the Elements}
\affil{Institute of Nuclear \& Particle Physics, Department of Physics \& Astronomy, Ohio University, Athens, Ohio 45701, USA}
\email{sl897812@ohio.edu, meisel@ohio.edu, browned@msu.edu}

\author[0000-0002-8403-8879]{Zach Meisel}
\altaffiliation{Affiliated with the Joint Institute for Nuclear Astrophysics—\\Center for the Evolution of the Elements}
\affil{Institute of Nuclear \& Particle Physics, Department of Physics \& Astronomy, Ohio University, Athens, Ohio 45701, USA}

\author[0000-0003-3806-5339]{Edward F. Brown}
\altaffiliation{Affiliated with the Joint Institute for Nuclear Astrophysics—\\Center for the Evolution of the Elements}
\affiliation{Department of Physics \& Astronomy, Michigan State University, East Lansing, Michigan 48824, USA}
\affiliation{Department of Computational Science, Mathematics, \& Engineering, Michigan State University, East Lansing, Michigan 48824, USA}

\begin{abstract}
Neutron stars cooling after sustained accretion outbursts provide unique information about the neutron star crust and underlying dense matter. 
Comparisons between astronomical observations of these cooling transients and model calculations of neutron star crust cooling have frequently been used to constrain neutron star properties such as the mass, radius, crust composition, and presence of nuclear pasta.
These calculations often use a fixed pressure at which the crust-core transition happens, though this quantity 
depends on
the dense matter equation of state.
We demonstrate that varying the crust-core transition pressure in a manner consistent with adopting various equations of state results in modest changes to the crust cooling light curve. This validates the approach adopted in most crust cooling studies to date, where the neutron star mass and radius are varied while leaving the crust-core transition pressure constant.
\end{abstract}


\section{Introduction} \label{sec:intro}
Neutron stars cooling after sustained accretion outbursts are a significant class of observables used to probe the structure of neutron stars and thereby the behavior of ultradense matter~\citep{Wijn17,Meis18,Baym18}.
The light curves from these cooling transients are often interpreted by modeling the thermal relaxation of the neutron star crust once accretion-driven heating ceases~\citep{Brow98,ushomirsky.rutledge:time-variable,Colpi:2001,Rutl02,Shte07,Brow09,Page13}. To date, a number of studies \citep[see, e.g.,][]{Dege14,Deib15,Merr16,Wate16,Meis17,Pari17,Oote18,Pari18b} have employed such model-observation comparisons to constrain several neutron star properties, such as the mass $M$, radius $R$, crust composition, accretion history, thermal structure, and presence of nuclear pasta.

A major feature of the accreted crust composition is the impurity, defined by the parameter $\Qimp \equiv n_{\rm ion}^{-1} \sum_j n_j (Z_j - \langle Z \rangle)^2$, where $Z_{j}$ is the nuclear charge of species $j$, with average $\langle Z\rangle$, and the number density of ions $n_{\rm ion}$ and species $n_{j}$, respectively. $\Qimp$ quantifies the thermal conductivity of the neutron star crust, which is dominated by electron-impurity scattering for the majority of the crust~\citep{Itoh93}. Generally, model-observation comparisons for cooling transients have determined that crusts are relatively pure, i.e. $\Qimp$ is relatively small. For example, $\Qimp\approx 3\textrm{--}4$ for XTE~J1701-462~\citep{Page13}, $4.4^{+2.2}_{-0.5}$ for KS 1731-26~\citep{Merr16}, $\lesssim6$ for MXB 1659-29~\citep{Pari18b}, and is $\sim 1$ for Swift J174805.3–244637~\citep{Dege15}, Aql~X1~\citep{Wate16}, 1RXS~J180408.9-342058~\citep{Pari17b}, and MAXI J0556-332 (though $\Qimp$ is not particularly relevant for such a hot crust; \citealt{Deib15}). Two significant exceptions (aside from cases considering substantial light-element enhancement in the neutron star ocean; \citealt{Medi14}) are EXO 0748-676 with $\Qimp=40$  \citep{Dege14}\footnote{\citet{Turl15} used $\Qimp=1$ to fit EXO 0748-676; however, those authors fixed the atmosphere temperature during accretion rather than letting the temperature be determined by accretion-related heating.}
and IGR J17480-2446 with $Q_{\rm imp}>25$~\citep{Vats18}\footnote{Reaction network calculations \citep{Lau18} find a larger $\Qimp$ in the accreted outer crust; in this region, however, electron-ion scattering dominates and impurity scattering is not as important \citep{Brow09}.}. 


However, $\Qimp$ determinations from crust cooling model-observation comparisons are sensitive to other assumptions about the neutron star crust. The time for heat to diffuse from a column depth $y\equiv \int_{r}^{\infty} \rho\,\dif r'$ to the neutron star surface is  
\citep{Heny69,Brow09}
\begin{equation}
\tau=\frac{1+z}{4}\left[\int^{y}_{0}\left(\frac{\CP}{\rho K}\right)^{1/2}\,\dif y'\right]^{2},
\end{equation}
where $1+z = 1/\sqrt{1-2GM/\left(Rc^{2}\right)}$ is the surface gravitational redshift, with $c$ being the speed of light and 
$G$ the gravitational constant. Here
 \CP\ is the specific heat per unit mass at constant pressure, $\rho$ is the mass density, and $K$ is the thermal conductivity. For the inner crust, $K\propto \Qimp$~\citep[see][and references therein]{Brow09} and \CP\ depends sensitively on the composition and whether or not the neutrons in the inner crust are paired \citep[see the discussion in][]{Meis18}. 
 $M$ and 
 $R$ enter into $\tau$ through the integral over column $y$ and through the redshift $1+z$.

Thermal gradients in the crust determine which direction heat diffuses. After accretion ceases, heat will flow from the location of accretion-powered heat sources to cooler regions at deeper and shallower depths. Predominant heating sources include the unknown source of shallow heating required to match observed cooling curves~\citep{Brow09,Deib15}, electron captures in the ocean and crust~\citep{Gupt07}, and deep crustal heating associated with neutron emissions and pycnonuclear fusion~\citep{Stei12}. As shallow heating tends to dwarf electron-capture heating, the latter is typically absorbed by the former in model calculations.

In reproducing observed transient light curves, which essentially requires reproducing $\tau(y)$, calculations often vary $M$, $R$, and $\Qimp$, along with properties mostly related to accretion. 
The crust-core transition pressure and the crust EOS is generally held fixed when fitting the lightcurve \citep[see, e.g.,][]{Brow09,Deib15,Merr16}, so that the crust thickness $\Delta r$ varies only through the dependence on $M$ and $R$. Upon expanding the Tolman-Oppenheimer-Volkoff (TOV) equation in $\Delta r/R$, one finds that 
to first order, $\Delta r$ is related to $M$ and $R$ by ~\citep{Zdun17,Sota17}
\begin{equation}
\label{e.crust-thickness}
    \Delta r \approx \chi\frac{R^{2}}{GM}\left(1-\frac{2GM}{Rc^2}\right),
\end{equation}
where $\chi = \int \dif P/(\rho + P/c^{2})$ depends only on the crust equation of state (EoS) and is integrated from the crust-core transition pressure to the photosphere. Equation~(\ref{e.crust-thickness}) is analogous to the Newtonian expression for the thickness of a thin atmosphere, $\Delta r \approx PR^{2}/(GM\rho)$. Since varying $M$ and $R$ independently is equivalent to changing the dense matter EoS, the factor $\chi$ should in principle also vary. Moreover, the thickness of the crust is sensitive to the assumed pressure of the crust-core transition, and many \citep[e.g.,][]{Stei15,Tsal19} have speculated that there should be a corresponding impact on the crust cooling.
In this work we explore the impact of the EoS modification to $\Delta r$ on light curves of cooling transient neutron stars.

In Section~\ref{sec:eos} we discuss the connection between $\Delta r$ and the EoS. We then model cooling transient light curves in Section~\ref{sec:dstar}, highlighting the individual impact of changes in $\Delta r$ for a variety of model assumptions. Section~\ref{sec:discuss} explains the insensitivity of model calculation results to changes in $\Delta r$ and  the implications for the extraction of neutron star properties from model-observation comparisons.

\section{The Equation of State and the Crust Thickness}\label{sec:eos}

The dense matter EoS provides the pressure-baryon density relation $P(n)$ needed to evaluate the TOV equations \citep{Oppe39,Tolm39} of general relativistic hydrostatic equilibrium for neutron star structure. At present, the EoS is insufficiently constrained, leading to a variety of predictions for neutron star properties \citep[see][for recent discussions]{Latt12,Latt16,Ozel16,Lali19}. 

\begin{table}[b!]
\centering
\caption{\label{tab:eos} Important EoS properties for this work.}
\begin{tabular}{l c c c c}%
\hline\hline
 & SLY4 & APR & BL & GM1 \\
\hline
$S$~[MeV]& 32.0  & 32.6 & 35.4 & 32.5 \\
$L$~[MeV] & 46.0 & 57.6 & 76.0 & 94.0 \\
$\ntr$~[fm$^{-3}$]& 0.0892 & 0.0807 & 0.0732 & 0.0577\\
$\log_{10}(P_{\rm t}/[{\rm g}~{\rm cm}^{-1}~{\rm s}^{-2}])$ & 32.9 & 32.8 & 32.7 & 32.5\\
$R_{1.4}$~[km] & 11.7 & 11.3 & 12.3 & 13.8 \\
$R_{2.0}$~[km] & 10.7 & 10.8 & 11.2 & 13.4 \\
\hline
\end{tabular}
\end{table}


Near nuclear saturation density $\nsat$, the energy per nucleon may be written \citep[see, e.g.,][]{Horowitz2014A-way-forward-i} as $\mathcal{E}(n,\alpha) = \mathcal{E}(n,\alpha=0) + S(n)\alpha^{2}$, where $\alpha = (n_{n}-n_{p})/(n_{n}+n_{p})$ is the neutron-proton asymmetry. Here $\mathcal{E}(n,\alpha=0)$ is the energy per nucleon of neutron-proton symmetric matter, and the quantity $S(n)$ is the nuclear symmetry energy, which is often expressed in an expansion about $\nsat$:
\begin{equation}
S(n) = J + \frac{n-\nsat}{3\nsat}L + \ldots.
\end{equation}
In this expression, $J\equiv S(n=\nsat)$ and
\begin{equation}
L = \left.3\nsat\left(\dd{S}{n}\right)\right|_{n=\nsat}.
\end{equation}
At $n=\nsat$, the energy of symmetric matter is minimized, i.e., $\left.\partial\mathcal{E}(n,\alpha=0)/\partial n\right|_{n=\nsat} = 0$; as a result, the pressure of pure neutron matter near saturation is $P(n,\alpha = 1) = n^{2}L/(3\nsat)$.

The EoS determines $\Delta r$ by setting the baryon density at which the crust-core transition occurs, $\ntr$. This density is approximately given by \citep{Newt13,Stei15}
\begin{equation}
\ntr = S_{30}(0.1327 - 0.0898 L_{70} + 0.0228 L_{70}^{2})\,\fermi^{-3},
\label{eqn:nt}
\end{equation}
where $S_{30}\equiv S/(\val{30}{\MeV})$ and $L_{70} \equiv L/(\val{70}{\MeV})$. This correlation is determined both by fits to nuclear experiments and theoretical calculations of pure neutron matter.
Using Equation~(\ref{eqn:nt}), the pressure at the base of the crust can be evaluated from an EoS as $P_{\rm t}\equiv P(\ntr)$. This quantity is more useful than $n_{\rm t}$ since $P$ is continuous throughout the crust and is therefore commonly used as a coordinate for depth. 

\begin{figure*}[htbp]
\centering
\includegraphics[width=\linewidth]{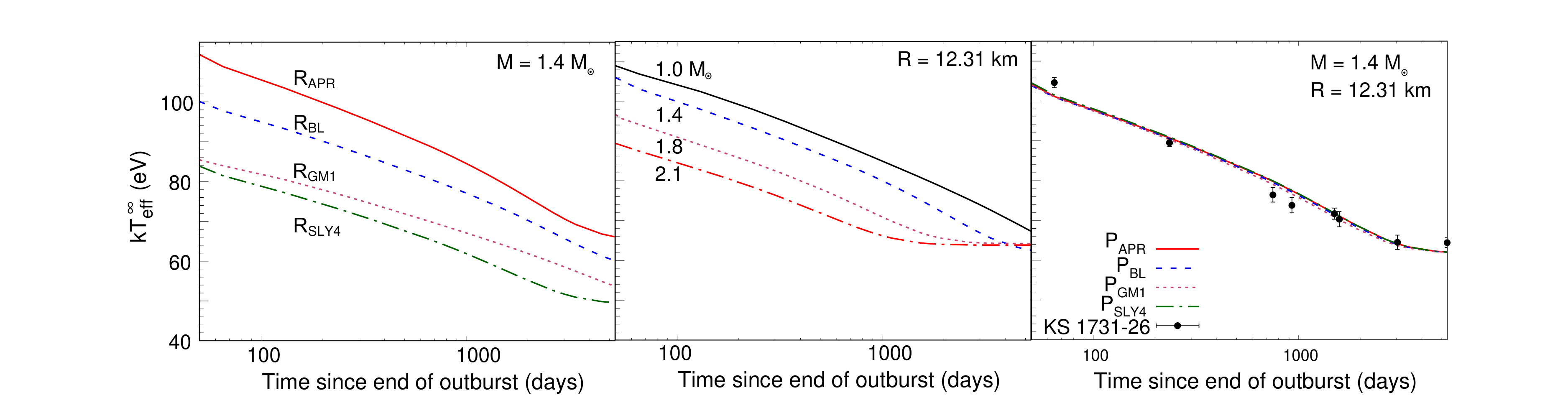}
\caption{ Impact of varied $R$, $M$, or $P_{\rm t}$ on crust cooling when the other two properties are fixed. The left and right panels use $R$  and $P_{\rm t}$, respectively, consistent with the EoS (see Table~\ref{tab:eos}). The left and center panels use an arbitrary fixed $P_{\rm t}$.
\label{fig:singleproperty}}
\end{figure*}

To explore the impact of the EoS on neutron star crust cooling, we have made a selection of nucleonic EoS that have a maximum neutron star mass $>2\,M_{\odot}$, so as to match observed pulsars \citep{Demorest2010A-two-solar-mas,Antoniadis2013A-Massive-Pulsa}. In order to sample some of the EoS phase space, we use the Skyrme EoS calculated with the SLY4 effective interaction~\citep{Gulm15,Chab98}, the microscopic EoS APR~\citep{Akma98}, the microscopic EoS BL~\citep{Bomb18}, and the relativistic mean field EoS GM1~\citep{Glen91}, where key quantities for this work are listed in Table~\ref{tab:eos}. The microscopic EoS are calculations using different nucleon-nucleon interactions and three-body forces. The latter three EoS have an inner crust described using SLY4~\citep{Douc01}. Note that the procedure for matching between this and the core EoS can impact $R$ and $\Delta r$~\citep{Fort16}. However, our purpose is to sample a variety of $\ntr\textrm{-}M\textrm{-}R$ combinations and so the exact $R$ and $\Delta r$ are not important. Furthermore, the impact of the crust EoS on $R$ and $\Delta r$ can be nearly eliminated using the recently developed method of \citet{Zdun17}.

\section{Crust Cooling Calculations}\label{sec:dstar}

Cooling transient light curves were calculated using the open-source code {\tt dStar}~\citep{Brow15}. {\tt dStar} models the thermal evolution of a neutron star crust after an extended accretion outburst by solving the general relativistic heat diffusion equation using the {\tt MESA}~\citep{Paxt11,Paxt13,Paxt15,Paxt18} numerical libraries. The microphysics is detailed in \citet{Brow09}. A number of thermodynamic, composition, structural, and numerical controls are available. Here we largely use a fixed set of conditions and explore the impact of modifying $P_{\rm t}$. 

Fixed quantities of interest for this work, inspired by \citet{Merr16} models of KS 1731-26, are the core temperature $T_{\rm c}=9.35\times10^{7}$~K, accretion outburst duration $\Delta t_{\rm out}=12.5$~yr, accretion rate $\dot{M}= 10^{17}$~g~s$^{-1}$, accretion-driven shallow heating $\mathcal{H}_{\rm sh}=1.36$~MeV$^{-1}$, shallow heating pressure boundaries $P_{\rm sh,low}=10^{27}$~g~cm$^{-1}$~s$^{-2}$ and $P_{\rm sh,hi}=10^{28}$~g~cm$^{-1}$~s$^{-2}$, low density boundary of the deep crustal heating $P_{\rm deep,low}=10^{30.42}$~g~cm$^{-1}$~s$^{-2}$, light element atmosphere column depth $y_{\rm lite}=10^{4}$~g~cm$^{-2}$, neutron star core mass $M_{\rm c} =1.4~M_{\odot}-M_{\rm crust}$ and radius (core radius plus crust thickness) $R=R_{\rm c}+\Delta r=12.31$~km, crust pressure boundaries $P_{\rm cr,top}=10^{27.2}$~g~cm$^{-1}$~s$^{-2}$ and $P_{\rm cr,bot}\equiv P_{\rm t}$, and crust impurity $\Qimp=4$. 
In general, $\Qimp$ varies throughout the crust \citep{Lau18}, but we choose a single value for the entire crust. This is partly to simplify the analysis, but the main justification is that the inner crust $\Qimp$ has the dominant impact on $\tau$ and it is unlikely to vary  substantially in this region (see Section~\ref{sec:intro}).

As changes in $P_{\rm t}$ largely change the depth that the crust extends down to, we also investigate the impact of modified $P_{\rm t}$ when making various assumptions about deep crustal heat release $\mathcal{H}_{\rm deep}$, the high-density boundary of the deep crustal heating $P_{\rm deep,hi}$ (here assumed to be equal to $P_{\rm t}$)\footnote{It is likely that deep heating is concentrated at densities just greater than neutron-drip~\citep{Zdun17}, but we choose to extend it to the base of the crust in order to maximize the potential impact of modifying $P_{\rm t}$.} 
, and the neutron superfluid gap, which ultimately determines $K$ near the base of the crust~\citep{Deib17}.

Figure~\ref{fig:singleproperty} shows example cooling curves, with data for KS 1731-26~\citep{Merr16} included for comparison, highlighting the impact of varying EoS related properties $M$, $R$, and $P_{\rm t}$. The X-ray flux is described by the effective temperature for an observer at infinite distance $k_{\rm B}T^{\infty}_{\rm eff}=(1+z)k_{\rm B}T_{\rm surf}$, where $k_{\rm B}$ is the Boltzmann constant and $T_{\rm surf}$ is the local surface temperature, as is customary for the corresponding observational data. The general trend is due to the heat deposited from nuclear processes in the deep crust reaching the surface at later times, relative to shallower layers, until crust-core equilibrium is achieved (e.g., at several thousand days in Figure~\ref{fig:singleproperty}). Smaller $R$ for fixed $M$ corresponds to larger $(1+z)$, stretching the cooling curve in time. The same is true for larger $M$ at a fixed $R$. A larger $P_{\rm t}$ implies a thicker $\Delta r$, which one would naively associate with a longer cooling time for a fixed $M,R$. We see, however, no such impact, which we explain in the following section.

\begin{figure}[htbp]
\centering
\includegraphics[width=\columnwidth]{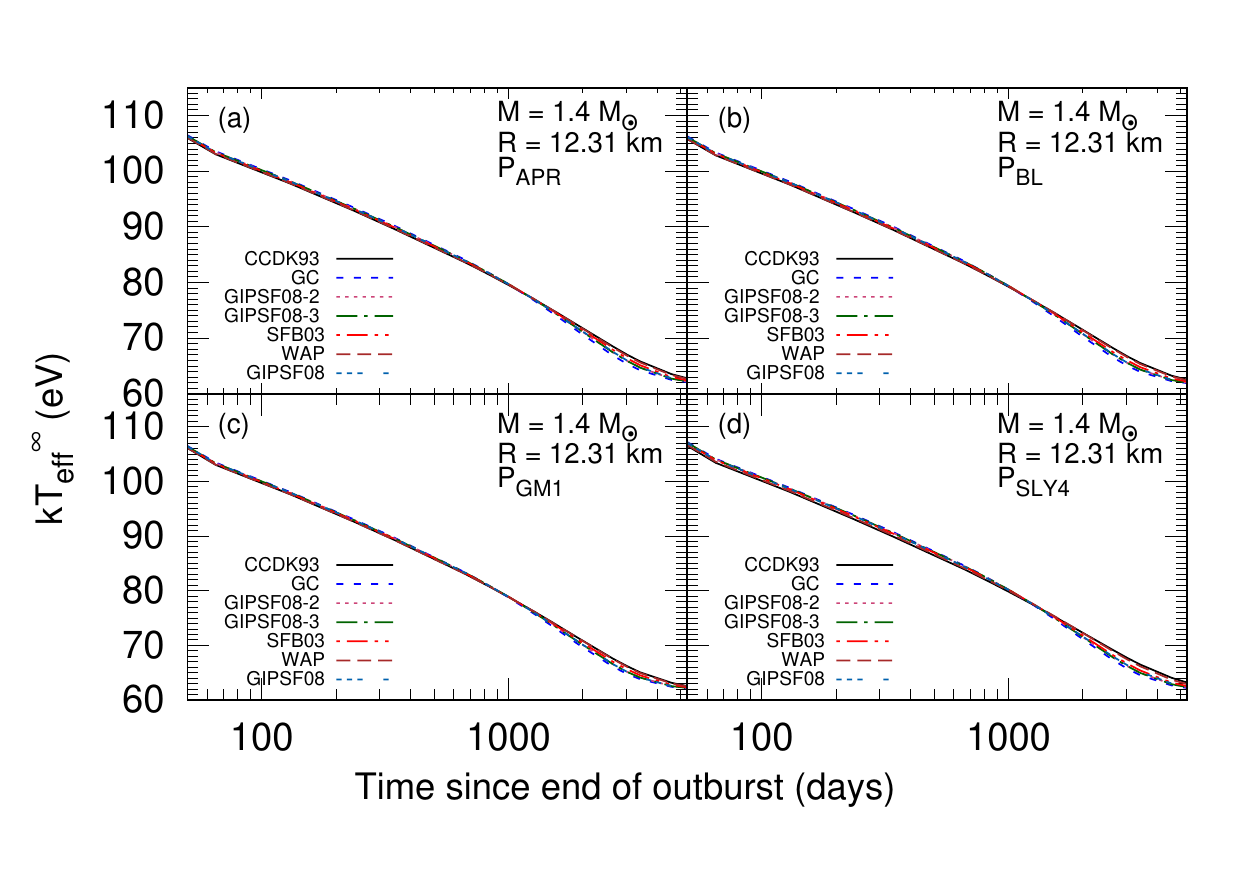}
\caption{Impact of varied neutron singlet pairing gaps on crust cooling when using $P_{\rm t}$ according to various EoS.
\label{fig:singlets}}
\end{figure}

\begin{figure}[htbp]
\centering
\includegraphics[width=0.8\columnwidth]{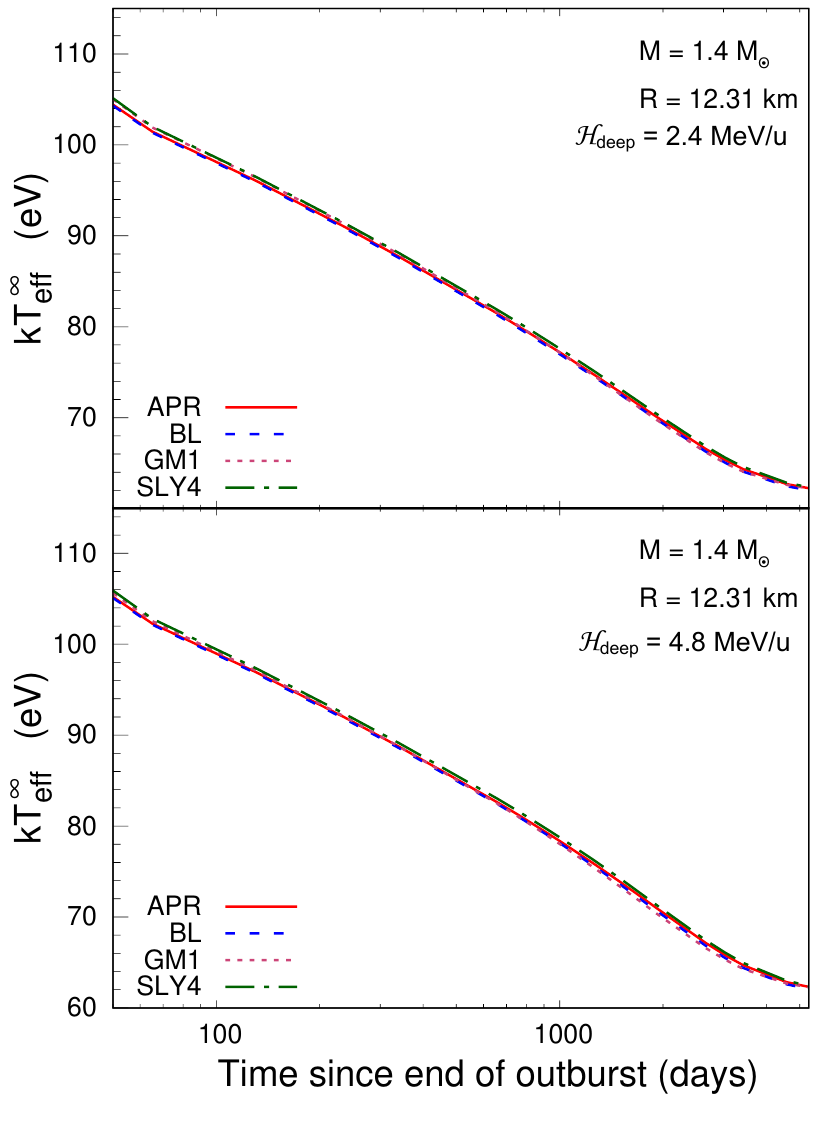}
\caption{Impact of varied $\mathcal{H}_{\rm deep}$ on crust cooling when using $P_{\rm t}$ according to various EoS.
\label{fig:heating}}
\end{figure}

To test if the insensitivity depended on assumptions of properties near the crust-core transition, we investigated the impact of varying $P_{\rm t}$ when adopting different $\mathcal{H}_{\rm deep}$ and models of the neutron singlet pairing gap. The impact of pairing gap models primarily relates to whether or not the neutron singlet pairing gap closes in the inner crust or in the core~\citep{Deib17}. Figure~\ref{fig:singlets} shows the impact of neutron singlet models CCDK93~\citep{Chen93}, GC~\citep{Geze08,Taka72}, GIPSF08~\citep{Gand08} with the gap closing at Fermi momentum $k_{\rm F}=1.1$~fm$^{-1}$, GIPSF08-2 at $k_{\rm F}=1.3$~fm$^{-1}$, GIPSF0-3  at $k_{\rm F}=1.5$~fm$^{-1}$, SFB03~\citep{Schw03}, and WAP~\citep{Wamb93}. Figure~\ref{fig:heating} shows the impact of varied $\mathcal{H}_{\rm deep}$, choosing the extreme cases featured in \citet{Stei12}.\\

\section{Discussion}\label{sec:discuss}

We find a negligible impact on the crust cooling light curve, despite $\Delta r$ changing nearly 20\% over the range of $P_{\rm t}$ explored here. This counterintuitive result can be understood by considering the information the crust cooling light curve communicates about the crust thermal structure and the return of the thermal structure to equilibrium after accretion turnoff.

\begin{figure}[htbp]
\includegraphics[width=\columnwidth]{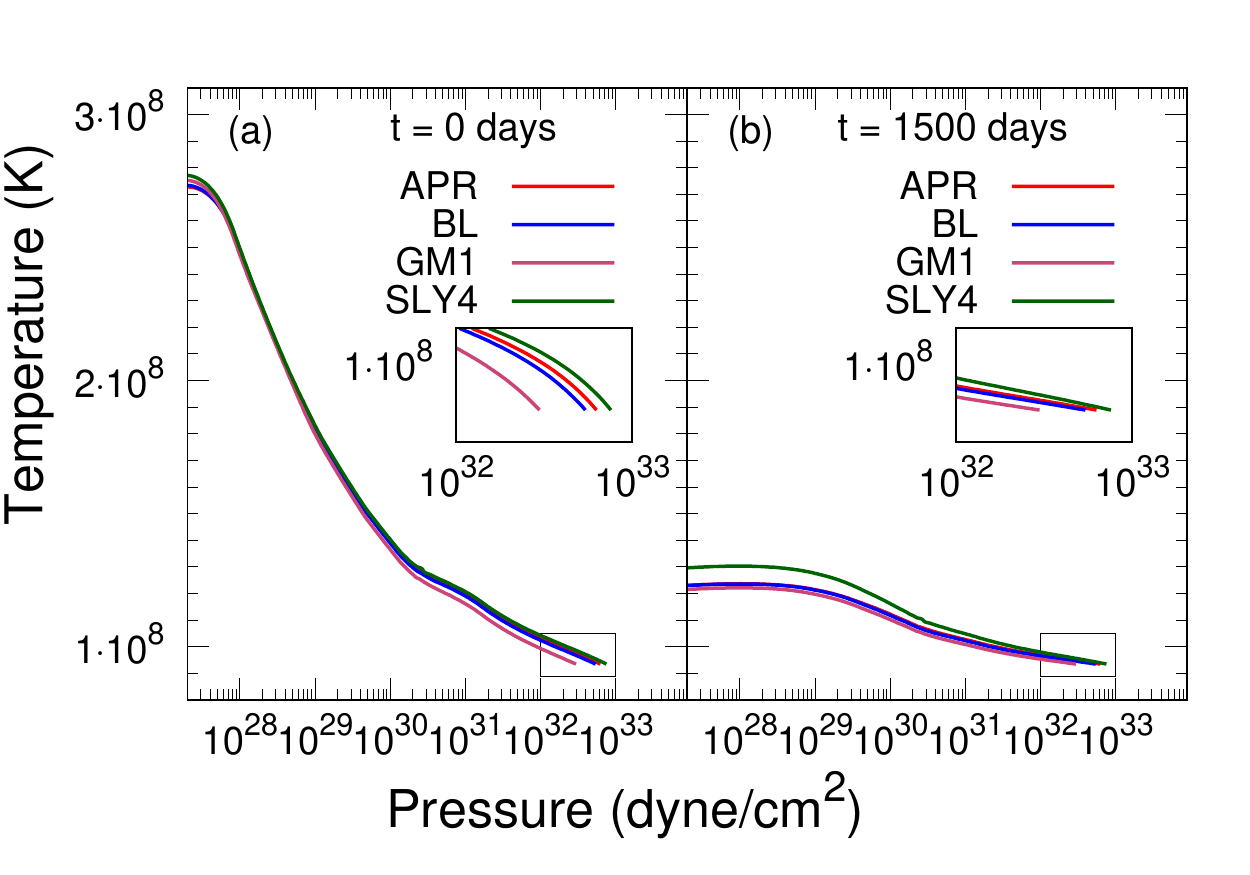}
\caption{Thermal profiles for {\tt dStar} calculations corresponding to the bottom panel of Figure~\ref{fig:heating} at accretion turnoff (a) and 1500~days later (b).
\label{fig:profile}}
\end{figure}

The X-ray luminosity emitted from the cooling transient source depends on the surface temperature at the time of emission. Sustained accretion results in a thermal structure that primarily decreases in temperature with increasing depth. As such, the surface cools to reach thermal equilibrium with continuously deeper depths as time progresses after accretion turnoff. Meanwhile, a large amount of heat diffuses into the relatively cold neutron star core. Therefore, by the time the cooling wave from the surface reaches the depths near $P_{\rm t}$, these regions have already nearly become isothermal with the core \citep[see, e.g.,][Fig.\ 3]{Page13}. This means that the extra crust thickness acquired from increasing $P_{\rm t}$ is essentially invisible. Figure~\ref{fig:profile} demonstrates this for the calculations presented in this work, featuring the thermal profiles for the case modeled in the bottom panel of Figure~\ref{fig:heating}, just after accretion turn-off and 1500~days into cooling. For the latter set of profiles, it is evident that the surface is in equilibrium with pressures much lower than $P_{\rm t}$, while the region near $P_{\rm t}$ is nearly indistinguishable regardless which EoS is adopted. 

\begin{figure}[htbp]
\includegraphics[width=\columnwidth]{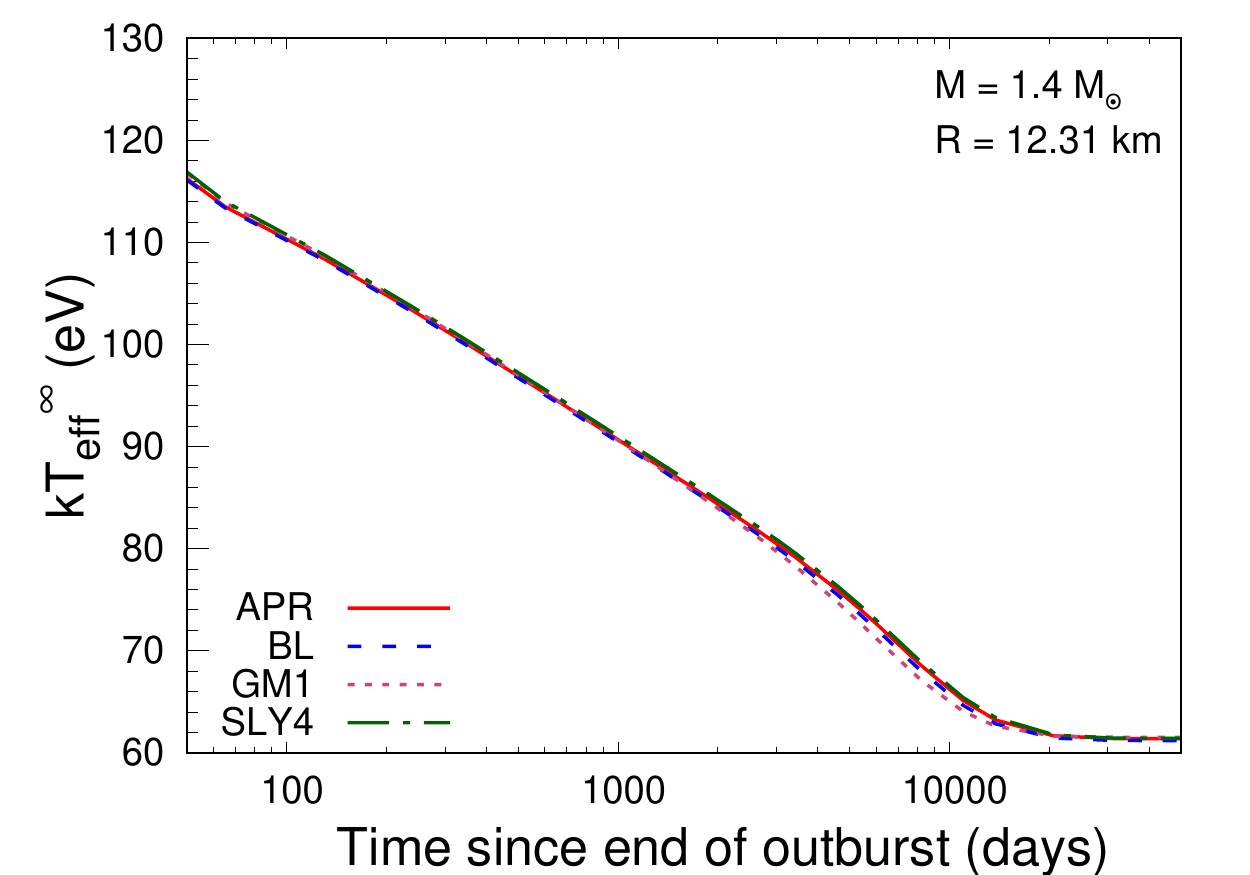}
\caption{Impact of adopting $P_{\rm t}$ from various EoS for $Q_{\rm imp}=25$. \label{fig:qimp}}
\end{figure}

For increased $Q_{\rm imp}$, the inner crust will take longer to cool into thermal equilibrium with its surroundings and therefore the thermal structure at accretion turn-off will be maintained longer. Figure~\ref{fig:qimp} demonstrates that sensitivity to $P_{\rm t}$ begins to set-in for $Q_{\rm imp}=25$, though differences between model results are still well within observational uncertainties. For context, $Q_{\rm imp}\approx20$ was found in the inner crust by \citet{Lau18} using crust reaction network calculations for an exceptionally hydrogen-rich X-ray bursting system. This was the largest $Q_{\rm imp}$ found in that work for the inner crust. 

The insensitivity to $P_{\rm t}$ significantly simplifies the task of modeling crust cooling for observed cooling transient sources. This is because $M$ and $R$ can be arbitrarily selected without the need to assume an EoS in order to consistently calculate $P_{\rm t}$. Additionally, the discrepancy in $P_{\rm t}$ between different procedures to match the EoS at the crust-core interface~\citep{Fort16,Gonz19} is unlikely to impact results for crust cooling models. Therefore, from this perspective, the use of a consistent EoS for the core and crust is not essential.

\begin{figure}[htbp]
\includegraphics[width=\columnwidth]{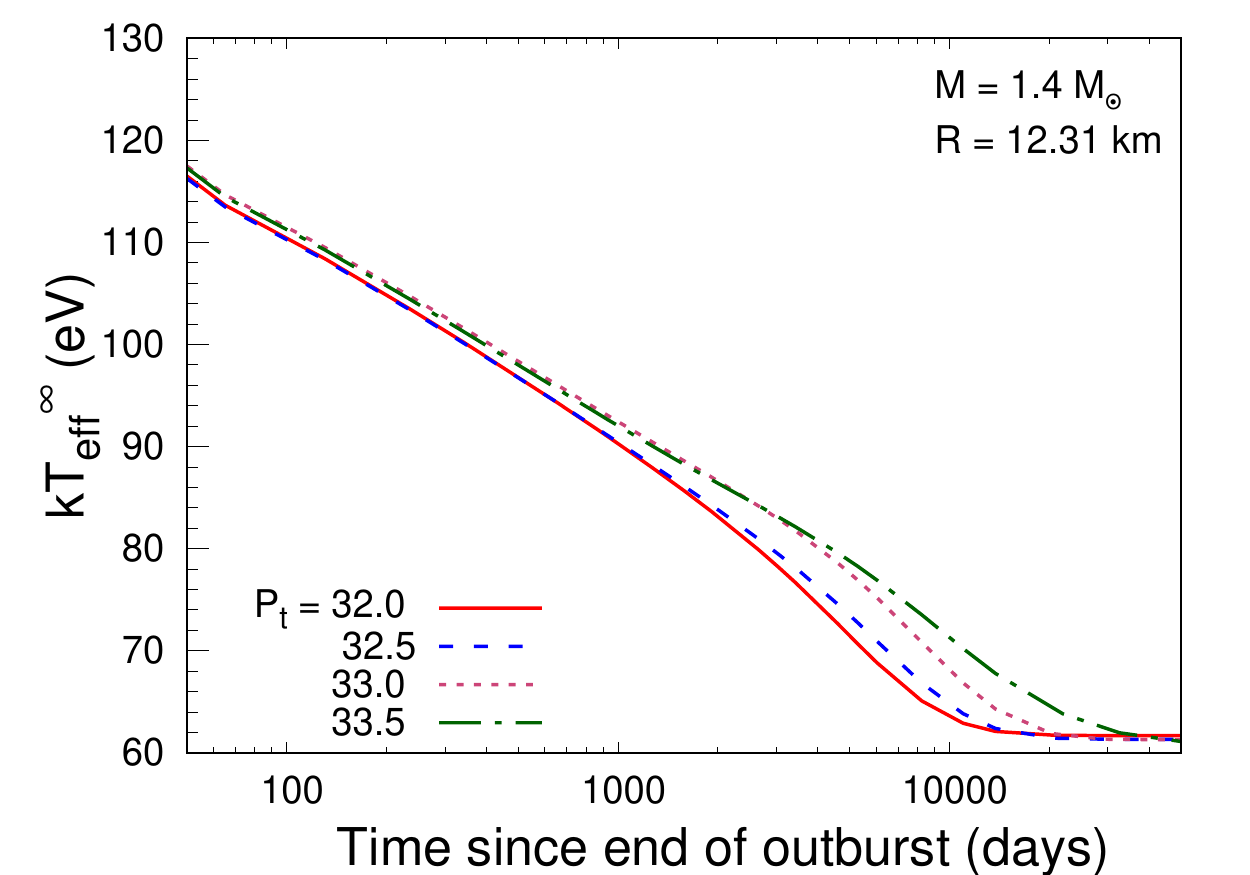}
\caption{Impact of adopting arbitrary $P_{\rm t}$ for $Q_{\rm imp}=25$. \label{fig:ArbitraryPt}}
\end{figure}

Thus far we have restricted ourselves to realistic EoS. One might wonder how far outside of the range in Table~\ref{tab:eos}
that $P_{\rm t}$ would have to be in order to result in an observable impact. This is addressed by Figure~\ref{fig:ArbitraryPt}. Relative to observational uncertainties (see Figure~\ref{fig:singleproperty}), a significant impact would require roughly 0.5~dex increase (decrease) above (below) the largest (smallest) $P_{\rm t}$ in Table~\ref{tab:eos}. 

\section{Conclusions}\label{sec:conclusions}

In summary, we investigated the impact of the $P_{\rm t}$ on the light curves of cooling transient sources as calculated via crust cooling models. Using {\tt dStar} models and conditions resembling those that reproduce observations of the source KS 1731-26, we show model results are insensitive to $P_{\rm t}$ when adopting pressures corresponding to realistic EoS. We find this is because the region near the crust-core interface reaches thermal equilibrium with the core long before the surface cools into equilibrium with these depths. This finding justifies the previously adopted approach in model-observation comparisons of neutron star crust cooling where $M$ and $R$ are varied irrespective of considering an EoS to determine a consistent $P_{\rm t}$. This also mitigates concerns about the dependence of $P_{\rm t}$ on the procedure used to match EoS at the crust and core interface.

\acknowledgments

We thank Ryan Connolly for useful discussions and {\tt CompOSE} (\url{https://compose.obspm.fr}) for providing EoS data.
This work was supported by the U.S. Department of Energy under grants DE-FG02-93ER-40756, DE-FG02-88ER40387, and DESC0019042.
 EFB is supported by the US National Science Foundation grant AST-1812838.
 We benefited from support by the National
Science Foundation under grant PHY-1430152 (Joint Institute
for Nuclear Astrophysics--Center for the Evolution
of the Elements).

\software{\dStar\ \citep{Brow15}}

\bibliographystyle{aasjournal}
\bibliography{References}

\end{document}